\documentclass[utf8]{frontiersSCNS} 

\usepackage{url,hyperref,lineno,microtype,subcaption}
\usepackage[onehalfspacing]{setspace}
\usepackage{ulem}

\def\keyFont{\fontsize{8}{11}\helveticabold }
\def\firstAuthorLast{Saur J.} 
\def\Authors{Joachim Saur\,$^{1,*}$}
\def\Address{$^{1}$Institute of Geophysics and Meteorology, University of Cologne, Cologne, Germany   }

\def\corrAuthor{J. Saur}
\def\corrEmail{saur@geo.uni-koeln.de}

\begin{document}
\onecolumn
\firstpage{1}

\title[Turbulence in Magnetospheres of Outer Planets]{Turbulence in the Magnetospheres of the Outer Planets} 

\author[\firstAuthorLast ]{\Authors} 
\address{} 
\correspondance{} 

\extraAuth{}

\maketitle

\begin{abstract}
  The magnetospheres of the outer planets  exhibit turbulent phenomena in an environment which is qualitatively different compared to the solar wind or the interstellar medium. The key differences are the finite sizes of the magnetospheres limited by their physical boundaries, the presence of a strong planetary background magnetic field and spatially very inhomogeneous plasma properties within the magnetospheres. Typical turbulent fluctuations possess amplitudes much smaller than the background field and are characterized by Alfv\'en times, which can be smaller than the non-linear interaction time scales. The magnetospheres of the outer planets are thus 
interesting laboratories of plasma turbulence. 
In Jupiter's and Saturn's magnetospheres, turbulence is well established thanks to the in-situ measurements by several spacecraft, in particular by the Galileo and Cassini orbiter. In contrast, the fluctuations in Uranus' and Neptune's magnetospheres 
are poorly understood due to the lack of sufficient data. 
Turbulence in the outer planets' magnetospheres have important effects on the systems as a whole. The dissipation of the turbulent fluctuations through wave-particle interaction is a significant heat source, which can explain the large magnetospheric plasma temperatures. 
%
%
 Similarly, turbulent wave fluctuations strongly contribute to the acceleration of particles responsible for the planet's auroras.

\tiny
 \keyFont{ \section{Keywords:} turbulence, outer planets, magnetosphere, weak turbulence, wave-particle interaction, aurora}
\end{abstract}

\vspace*{1cm}

\section{Introduction and origin of turbulence}

All four outer planets of the solar system, Jupiter, Saturn, Uranus and Neptune, possess strong internal dynamo magnetic fields with polar field strength of $\sim$ 2,000,000 nT,  40,000 nT, 100,000 nT and 90,000 nT, respectively 
%
%
\citep{conn07,ness10}. 
They are also fast rotators with periods of 9.9 hrs, 10.7 hrs, 17.2 hrs and 16.1 hrs, respectively \citep{seid02}. These two effects lead to large and rotationally dominated magnetospheres with average magnetopause standoff distances on the sub-solar side of  65, 20, 20, and 26 planetary radii, respectively \citep{vasy09,bage09,bage13}.

One of the most important difference between the magnetospheres of the outer planets and the magnetosphere of Earth is the presence of large mass sources well inside the outer planets' magnetospheres and the resultant effects generated by the associated radial mass transport.
In case of Jupiter, its inner Galilean moon Io, located deeply within the magnetosphere at 6 R$_J$, produces about 10$^3$ kg s$^{-1}$ of SO$_2$, which eventually gets ionized \citep{thom04}. An additional, but smaller source of mass is its moon Europa with about 50 kg s$^{-1}$ of H$_2$O \citep{bage20}. In case of Saturn, the mass source is the moon Enceladus, with a time-variable injection rate of 200-1000 kg s$^{-1}$ of H$_2$O \citep{doug09,saur11,fles10a,hedm13}. In case of Uranus and Neptune, also major moons orbit within their magnetosphere. However, their mass loss rates are not well constrained. They are considered smaller compared to those in the magnetospheres of Jupiter and Saturn based on the significantly smaller mass densities in the magnetospheres of Uranus and Neptune \citep[e.g.,][]{bage09,bage13}.

 \begin{figure}[h!]
\begin{center}
\includegraphics[width=20cm]{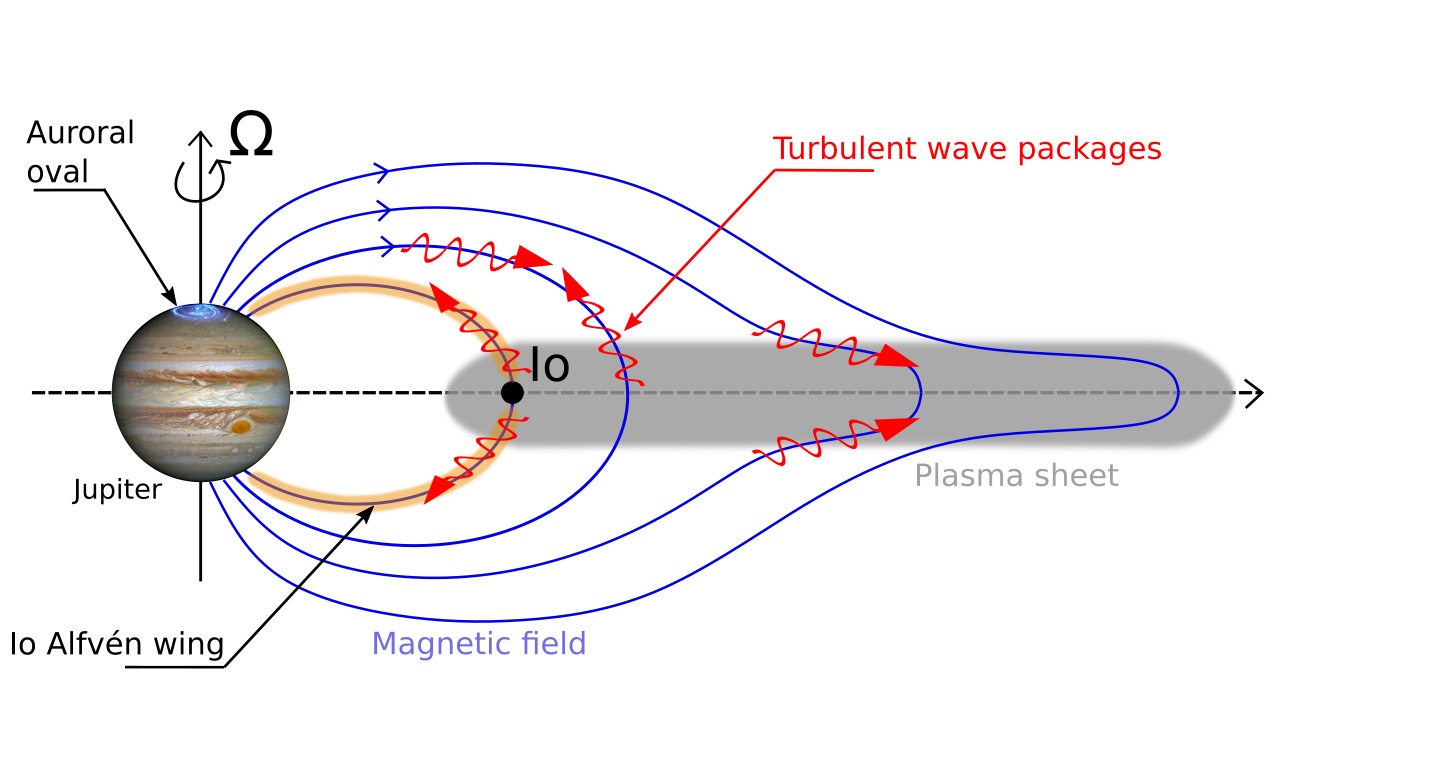}
\end{center}
\caption{Sketch of turbulent wave packages in Jupiter's magnetosphere. The interaction of counter-propagating Alfv\'en wave packages generates a turbulent cascade.
Waves packages are reflected at the ionosphere of Jupiter or other density gradients in the magnetosphere.
}
\label{fig:sketch}
\end{figure}
 The centrifugal force in the fast rotating magnetospheres causes the plasma to be concentrated near the equatorial regions (called plasma sheets, see Figure \ref{fig:sketch}) and drives transport of plasma radially outward. Due to conservation of angular momentum the outward moving plasma does not to fully corotate with the planet. This generates two types of magnetospheric stresses: Stretched magnetic field lines in the radial direction  from the radial transport (see Figure \ref{fig:sketch})  and  bend backed magnetic field lines in azimuthal direction due to the subcorotating plasma. These magnetic stresses couple to the planets' ionospheres and cause transport of angular momentum between the planets' ionospheres and magnetospheres \citep{hill01,gold07}. The stress balance is however not in steady state, but constantly disturbed due to non-continuous radial transport of plasma, which is observed to occur through the interchange of mass loaded magnetic flux tubes with adjacent less loaded flux tubes further out \citep{kive97c}. The imbalance of magnetic stresses cause Alfv\'en waves to propagate along the magnetic field to achieve stress balance. The Alfv\'en wave packages are partially reflected at the ionospheres or other boundaries, such as large density gradients of the plasma sheet. The resulting counter-propagating Alfv\'en wave packages interact nonlinearly and thus generate a turbulent cascade \citep{saur02,saur18a}. 
On the MHD scales the magnetospheric plasma sheet is thought to be a key region where the cascade is being driven \citep{saur02}.

The non-linear interaction of counter-propagating wave packages is the fundamental building block of MHD and plasma turbulence. It is established  observationally, theoretically and in numerical modeling \citep[e.g.,][]{gold95,howe13}.  There are several key differences between turbulence in the solar wind \citep[e.g.,][]{brun13}  and the turbulence in the magnetospheres of the outer planets. The magnetospheric fluctuations are small amplitude fluctuations $\delta B \ll B_0$ compared to the dominant planetary background magnetic field $B_0$. The magnetospheres are also  finite in size. This implies that turbulent scales are bound and a maximum length scale for the inertial range of turbulence exists.
It also causes reflection of the wave packages at the ionospheric boundaries. Additionally the magnetospheres are highly inhomogeneous with large densities and small magnetic field strength in the magnetospheric plasma sheets and vice versa at high latitudes (see Figure \ref{fig:sketch}). Therefore the generator region, i.e. where the turbulent cascade is driven, and the dissipation regions of the turbulent fluctuations do not need to be spatially collocated \citep[]{saur18a}.
Another interesting difference between the solar wind and the magnetospheres of the outer planets is that the composition of ions has a much larger spread in mass. The solar wind is mostly composed of H$^+$ and He$^{++}$ compared to H$^+$, O$^+$, S$^+$, H$_2$O$^+$ ... in Jupiter's or Saturn's magnetospheres. This causes a larger spread of gyro frequencies and inertial length scales and will influence the pathways of turbulent dissipations.  

In the reminder of this review we will present observations of turbulent fluctuations in Jupiter's and Saturn's magnetospheres. Additionally, we discuss their impact on overall properties of these magnetospheres such as aurora or temperature structure.

\vspace*{1cm}

\section{Magnetospheric Turbulence}
Turbulence and its implications for the magnetospheres of Jupiter and Saturn have been studied by a number of authors. 
There are no published studies about turbulence in the magnetospheres of Uranus and Neptune due to a lack of appropriate data. 

\vspace*{1cm}

\subsection{Jupiter magnetosphere}
Turbulence in Jupiter's magnetosphere has been investigated on various levels of detail.
Before the Galileo epoch, \citet{nish76} considered whistler mode turbulence 
for pitch angle diffusion of energetic electrons and ions. \citet{barb81a} and \citet{barb84}  investigated ion heating due to high-frequency 
kilohertz plasma wave turbulence and low frequency Alfv\'enic turbulence, respectively. 
\citet{glas95} suggests a Kolmogorov-type of turbulence 
in the very low-frequency milli-Hertz range based on Voyager magnetic field measurements.

Magnetic field measurements by the Galileo orbiter led to a dedicated turbulence analysis by \citet{saur02}. The analysis covered the magnetic field fluctuations in Jupiter's middle magnetosphere (9-24 R$_J$) within the magnetohydrodynamic (MHD) range between  2.8 $\times$ 10$^{-4}$ -- 3.6 $\times$ 10$^{-2}$ Hz, i.e., below the ion cyclotron frequency. Even though the relative velocity between the magnetospheric plasma and the spacecraft is much smaller compared to spacecraft in the solar wind, the Taylor theorem can still be applied \citep{saur13}. The properties of the magnetic field fluctuations in Jupiter's middle magnetosphere differ from those in the solar wind. In the Jovian system, the ratio of the magnetic fluctuations $\delta B$ compared to the background magnetic field $B_0$ is less than 10$^{-1}$, while it is of order unity in the solar wind.  
A weak turbulence cascade is expected in Jupiter's middle magnetosphere based on the ratio $\epsilon =\tau_A / \tau_{nl}$ between the Alfv\'en time $\tau_A$ and the non-linear interaction time $\tau_{nl}$. This ratio $\epsilon$ lies in the range of 0.06 to 0.34 and is indicative of weak turbulence \citep{ng97,galt00}.  Magnetic field fluctuations parallel to the background magnetic field exhibit a power law with
spectral index of -2 \citep{saur02} also indicative of weak turbulence \citep{ng97,galt00}. The other components often show deviations from clear power laws possibly due to other perturbations in the Jupiter's highly dynamic magnetosphere or locally non-fully developed turbulent cascades.

Turbulence in Jupiter's magnetosphere has subsequently been analyzed by \citet{tao15} who combined low and high-time resolution magnetic field data of the Galileo spacecraft within MHD and kinetic scales, i.e., within 1 $\times$ 10$^{-4}$ Hz to 1 Hz. The resultant magnetic field spectra show spectral breaks well organized by the ion scales (i.e., ion cyclotron frequency, the Doppler-shifted ion inertial length scales and the ion gyroradius, which assume similar values in Jupiter's middle magnetosphere). For frequencies smaller than those associated with the ion scales, the spectral indices lie in the range of -0.6 and -1.9 and for higher frequencies within the range of -1.7 and -2.5 \citep{tao15}. The authors also show that the turbulence is intermittent, in particular in the equatorial region of Jupiter's magnetospheric plasma.

\vspace*{1cm}

\subsection{Saturn magnetosphere}

\begin{figure}[h!]
\begin{center}
\includegraphics[width=15cm]{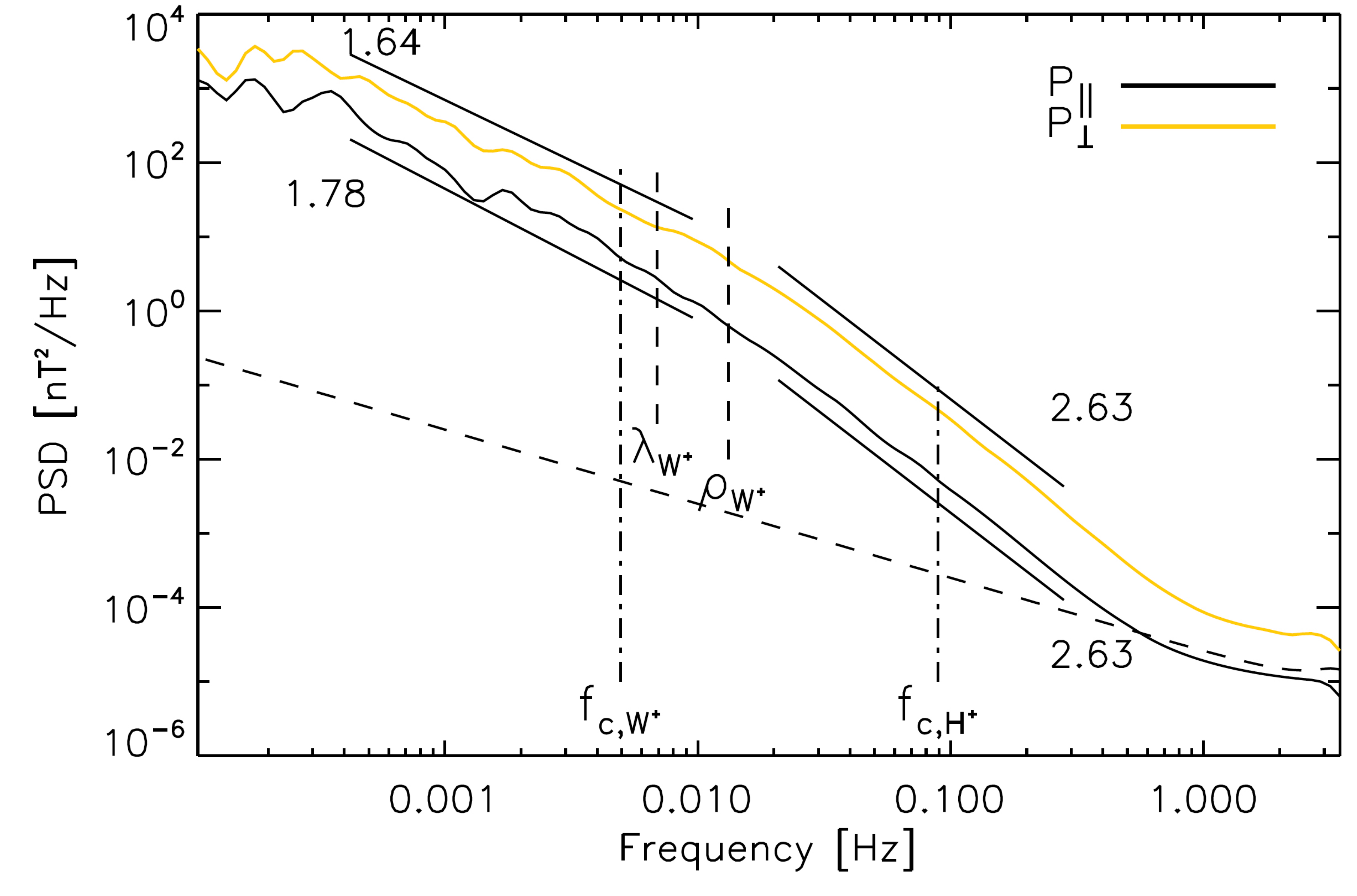}
\end{center}
\caption{
%
%
 Power spectral density (PSD) of turbulent magnetic field fluctuations in Saturn's magnetosphere (from \citet{vonp14}). $P_\perp$ and $P_\parallel$ refer to the power spectral densities of fluctuations perpendicular and parallel to the background magnetic field, respectively. $\lambda_{W^+}$ and
$\rho_{W^+}$indicate the Doppler shifted  ion inertial length scale and gyroradius of water ions, respectively, and 
$f_{c, W^+}$  and $f_{c,H^+}$ are the cyclotron frequencies of water ions and protons, respectively.}
\label{fig:spectrum}
\end{figure}

Saturn's magnetosphere also hosts small scale turbulent magnetic field fluctuations as demonstrated by \citet{vonp14}. The average amplitudes of the fluctuations compared to the magnetospheric background field assumes values of $\delta B / B$ = 0.07. An exemplary spectrum is shown in Figure \ref{fig:spectrum}.  On MHD scales the spectral slope of the fluctuations perpendicular and parallel to the background magnetic field varies between -0.8
 and -1.7. 
At ion scales, i.e., near the ion inertial length scale or the Doppler-shifted ion gyroradius, a spectral break occurs, with a sub-ion, i.e., high-frequency, slope of -2.3 inside of radial distances of 9 $R_S$ (see Figure \ref{fig:spectrum}). Further outside in the magnetosphere the sub-ion spectrum steepens to -2.6. These slopes could be consistent with turbulence of kinetic Alfv\'en waves. The probability density function of the fluctuations have non-Gaussian tails and a power law increase of the flatness indicates that the turbulence is intermittent \citep{vonp14}.  The turbulence on the kinetic scales is estimated to be strong in Saturn's magnetosphere with $\epsilon > 1$, i.e., the non-linear time is shorter than the wave propagation time, which could result from Saturn's weaker planetary magnetic field compared to Jupiter's field.

In a follow-up study,  \citet{vonp16} investigated the spatial and temporal structure of the turbulence in Saturn's magnetosphere. In local time coordinates, enhanced fluctuations are seen at noon, possibly resulting from flux tube interchanges \citep[e.g.,][]{kive97c}, which are considered a prime source of the turbulence. In a frame rotating with Saturn, increased fluctuations are seen at 65$^{\circ}$ southern and 250$^{\circ}$ northern magnetic phase. The later correlation is enigmatic and is related to the still unresolved planetary period oscillations of Saturn \citep{espi00,prov09,gurn09,vonp16}. The oscillations might result from a coupling between Saturn's atmosphere, ionosphere and magnetosphere. The variability in the turbulence can thus be considered as an effect of the Alfv\'enic energy fluxes underlying these coupling processes. The spatial variability of the turbulent fluctuations in Saturn's magnetosphere has been subsequently studied by  \citet{kami17}. The authors also found enhanced fluctuations between 10 hr and 20 hr local time and the turbulence to be more quiet between 3 hr and 9 hr local time.

\vspace*{1cm}

\subsection{Large scale implications of turbulence}

In addition to serving as a laboratory for turbulence studies, the fluctuations in the outer planets' magnetospheres have major implications on overall properties of their planetary systems. 

\vspace*{1cm}

\subsubsection{Magnetospheric heating} 

The magnetospheres of Jupiter and Saturn possess ion temperatures which strongly increase   
from the plasma source regions radially outwards by approximately two orders of magnitude
to values up to nearly 10$^8$ K \citep{bage11}.
These observations are considered a long standing puzzle since major energy input rates are required to explain these temperature increases. Otherwise the radial transport of magnetospheric plasma would lead to adiabatic cooling with distance.
Dissipation of turbulent magnetic field fluctuations is a very powerful energy source which can explain the temperature increase.

First suggestions of the importance of turbulent heating in Jupiter's magnetosphere in addition to pickup energization and subsequent radial transport date back to \citet{barb84}. Based on the observed spatial distribution of the magnetic field fluctuations and a model for a weak turbulent cascade, \citet{saur04a} calculated the dissipation rate in Jupiter's middle magnetosphere to a total amount of 5 $\times $ 10$^{12}$ W 
and demonstrated that this rate and its spatial distribution can explain the observed radial temperature profile 
\citep{fran02a}. To a similar conclusion came  \citet{ng18} who used a radial transport model based on advection  and not on flux tube diffusion as in \citet{saur04a}. The turbulent dissipation rates are consistent with the required heating rate independently extracted from Galileo spacecraft  observations by \citet{bage11}.  A possible dissipation mode could be ion cyclotron damping  in particular of the heavy sulfur and oxygen ions of Jupiter's magnetosphere \citep{saur18a}. 
In Saturn's magnetosphere the dissipation of the magnetic field fluctuations based on a strong turbulence model could provide power
 on the order of 10$^{11}$ W \citep{vonp16,kami17,neup21}, which is roughly consistent with the energy needed to heat the magnetosphere to its observed temperatures \citep{bage11}.

\vspace*{1cm}

\subsubsection{Aurora}

Jupiter's main auroral oval lies on magnetic field lines that map to an equatorial distance of about 20 - 30 R$_J$ \citep[e.g.,][]{clar98,clar02,hill01}
The turbulent power of the magnetic fluctuations in Jupiter's magnetosphere maximizes at the same radial distance \citep{saur03}. This  distance is also the region where the corotation of Jupiter's magnetosphere breaks down and magnetospheric-ionospheric  coupling currents maximize 
as derived by 
\cite{hill79,hill01,cowl01}. The turbulent fluctuations result from the non-steady radial transport and the resultant imbalanced stresses between the magnetosphere and Jupiter's ionosphere. The stress balance is achieved by Alfv\'en waves propagating between both regions. When the Alfv\'en waves reach the polar regions of the magnetosphere, the perpendicular length scales of the Alfv\'en waves grow small due to Jupiter's increasing magnetic field. Just above the ionosphere, the electron inertial length scale assumes its large values of $\sim$50 km due to the low electron density. When the turbulent cascade reaches this scale, the fluctuations are subject to electron Landau damping resulting in stochastically accelerated electrons \citep{saur18a}. This process is consistent with the bi-directional and energetically broad band electron distributions  recently measured by the Juno spacecraft \citep[e.g.,][]{mauk17,clar18}. These electrons are one of the key contributors to Jupiter's main auroral oval. Magnetic field fluctuations measured by the Juno spacecraft near the auroral particle acceleration region showed a spectrum with a slope of -2.29 $\pm$0.09 within 0.5 -- 5 $\times$ 10$^{-1}$ Hz \citep{gers19}. They could be due to kinetic/inertial Alfv\'en waves, however, the underlying turbulent processes for this spectrum are not fully clear yet as the Alfv\'en velocity is close to the speed of light in this region. The aurora of Saturn is less well understood, but also within Saturn's magnetosphere bi-directional energetically broad band electron beams have been observed on field lines which are connected to the aurora \citep{saur06,mich09} and which appear to be driven by stochastic, turbulent processes \citep{vonp16}.

\vspace*{1cm}

\subsection{Turbulence in Alfv\'en wings of the Moons}
A new and evolving area of turbulence research in the outer planets' magnetosphere are the Alfv\'en wings generated by the moons. In Figure \ref{fig:sketch}, we exemplarily show Jupiter's moon Io. The moon is an obstacle to the rotating magnetospheric plasma and  generates Alfv\'en waves, which propagate along the rotating background magnetic field towards Jupiter. In the rest frame of each moon two standing waves develop (north and south, respectively). 
They are referred to as Alfv\'en wings \citep{neub80,neub98,goer80,saur13}. The moons can thus be considered as gigantic Alfv\'en wave antenna. The Alfv\'en wings  are reflected at gradients of the background density and generate filamented structures \citep{chus05,hess11}. The magnetic field fluctuations in Io's Alfv\'en wings at high latitudes have been recently measured by the Juno spacecraft. Their spectrum is consistent with a power law spectral index of -2.35 $\pm$0.09 within 0.1 -- 800 Hz \citep{sula20}. 
When the filamented waves approach kinetic scales, wave-particle interaction sets in and can energize electrons through electron Landau damping and ions through cyclotron damping \citep{hess10,szal18,szal20,saur18a}. Similar particle acceleration processes have also been observed in the Alfv\'en wings of Europa \citep{alle20} and Ganymede \citep{szal20a}.

\vspace*{1cm}

\subsection{Magnetosheaths' of Jupiter and Saturn}
Turbulence in the magnetosheaths of the outer planets has received very little attention to date. Turbulence in the magnetosheaths is distinct from turbulence within the outer planets' magnetospheres. The magnetosheaths are characterized by large amplitude magnetic fluctuations $\delta B/ B_0 \sim 1$ \citep{hadi15}. The ultimate source of free energy stems from the shocked solar wind plasma with Alfv\'en Mach numbers as high as $\sim$100 \citep{mast13},  which causes enhanced plasma temperatures and large plasma beta turbulence. 

\cite{alex08} analyzed turbulent magnetic field fluctuations in Saturn's magnetosheath obtained the Cassini spacecraft and identified  Alfv\'en vortices similar to those found in Earth's magnetosheath. 
Alfv\'en vortices are non-linear magnetic structures associated with sheared velocity perturbations propagating obliquely to the external field.  \citep{hadi15} also studied Saturn's magnetosheath and found power-spectral energy densities of the magnetic field fluctuations with a $\sim$$f^{-1}$ scaling at MHD scales and an $\sim$$f^{-2.6}$ scaling at sub-ion scales.  The authors also found based on the compressibility of magnetic and density fluctuations that compressible magnetosonic  slow modes at MHD scales dominate rather than the Alfv\'en mode. Magnetic field turbulence in Jupiter's magnetosheath was studied by \cite{bolz14} based on Ulysses measurements. \cite{bolz14} considered a similar frequency range as \cite{hadi15}, but fitted the power spectral density to only one power law instead of two and thus found an average slope of around $-2$. 

\vspace*{1cm}

\section{Discussion: Outstanding issues and outlook}

For a more comprehensive understanding of turbulence in the outer planets' magnetospheres further measurements and studies are required. The Galileo spacecraft recorded only a small fraction of its mission in a high-frequency modus. The Juno spacecraft is currently exploring the high latitude regions of Jupiter's magnetosphere including the moons' Alfv\'en wings. With its instrumentation and its polar orbit detailed studies of turbulence and the associated wave-particle processes of turbulent dissipation are on their way. 

New missions with appropriate instrumentation to Uranus and Neptune are being discussed \citep[e.g.,][]{arri14}, but no mission has been decided upon. 
The orientations of the planetary magnetic moments at Uranus and Neptune are highly tilted compared to the planets' spin axis. Also the plasma densities and the magnetic field fluctuations in these magnetospheres are much smaller compared to those of Jupiter and Saturn. It will be interesting to see how these properties influences the possible turbulent nature of the field and plasma fluctuations.

Further studies are warranted on where and what types of turbulent cascades are driven in the magnetospheres of the outer planets. For example, are there cascades on kinetic scales outside the plasma sheets? Also a better understanding of the nature of the dissipation mechanisms of the turbulent fluctuations and the resultant ion or electron distribution functions is an important matter. Further observational studies might aim to better resolve the effects of the various dissipation and turbulent break scales expected to occur at the inertial lengths scales and the gyroradii of the electrons and the various ion species, respectively. Because the magnetospheres are highly inhomogeneous, the spatial variability of these parameters and their effects can be more easily studied compared to the solar wind. 
Suggestions for the roles of Landau and ion cyclotron damping of Alfv\'enic turbulence in the magnetosphere of Jupiter have been made \citep[e.g.,][]{saur18a}. However, the role of other wave modes or non-resonant wave-particle interaction mechanisms warrant further analysis.  

Numerical simulations with hybrid or particle-in-cell models can be a useful tool to better understand the dissipation mechanisms and the associated transport of particles and energy \citep[e.g.,][]{dela21}. Future models could focus on more 
realistic magnetic field geometries with current sheets and inhomogeneous plasma densities along field lines.

Turbulence in the magnetosheaths of the outer planets have not received much attention. Dedicated  study might investigate the structure of turbulence downstream of the planets' bow shocks to test the universality of turbulent evolution in Earth's and other planets' magnetosheaths. In the much larger magnetosheath of the outer planets' turbulence might evolve toward a more fully developed turbulent states compared to Earth. 
While turbulence in the middle magnetospheres of Jupiter and Saturn is unique due to the plasma of the moons and the fast rotating planets, the existence of turbulence in the magnetospheric tails and the magnetosheaths could be universal across all magnetized planets.

Turbulent fluctuations might  be used as a diagnostic tool to probe the interior of the icy moons in the outer solar system. Icy moons, such as Europa or Ganymede, possess subsurface water oceans \citep{kive00,kive02,saur10,saur15}. These saline and thus electrically conductive oceans have been detected through electromagnetic induction caused by time-variable  magnetospheric fields. For existing observations only the $\sim$10 hr rotation period of Jupiter could be applied. Turbulent fluctuations provide a broad range of frequencies which can be used in future induction studies to additionally probe the interior of the moons  (e.g., with JUICE or Europa Clipper data).


In summary, compared to the solar wind or the interstellar medium, the magnetospheres of the outer planets are complementary laboratories to study plasma turbulence.
These magnetospheres are distinct due the planets' large background magnetic fields, the bound and inhomogeneous nature of the turbulent systems, where the location of the turbulent cascade and the turbulent dissipation are not necessarily collocated.



\vspace*{1cm}

\section*{Conflict of Interest Statement}

The author declares that the research was conducted in the absence of any commercial or financial relationships that could be construed as a potential conflict of interest.

\section*{Author Contributions}

J.S. selected the material for this review and wrote the manuscript.

\section*{Funding}
This project has received funding from the European Research Council
(ERC) under the European Union’s Horizon 2020 research and innovation
programme (grant agreement No. 884711).

\section*{Acknowledgments}
J.S. thanks Sascha Janser for helpful comments on the manuscript  and Filip Elekes for the graphics support.
\bibliographystyle{frontiersinSCNS_ENG_HUMS} 

\end{document}